\documentstyle{article}
\title{Crystalizing the Spinon Basis}
\author{Atsushi Nakayashiki\
and Yasuhiko Yamada\\
Graduate School of Mathematics\\
Kyushu University}
\date{}

\begin{document}
\def\uq{U_q(\widehat{sl_2})}
\def\be{{\bf \epsilon}}
\def\la{\lambda}
\def\La{\Lambda}
\def\c{\varphi^\ast}
\def\v{\varphi}
\def\ot{\otimes}
\def\bz{{\bf Z}}
\def\et{\tilde{e}}
\def\ft{\tilde{f}}
\def\jt{\tilde{j}}
\def\ep{\epsilon}
\def\ca{{\cal A}}
\def\pek{{\cal P}^k}
\def\enpn{\epsilon(j_{n+1}+j_n)}
\def\ennm{\epsilon(j_n+j_{n-1})}
\def\knpn{K(p_{n+1},p_n)}
\def\knnm{K(p_n,p_{n-1})}
\def\cb{{\cal B}}
\def\br{r}
\def\th{\tilde{H}}
\def\cp{{\cal P}}
\def\bp{\bar{p}}
\def\ra{\rightarrow}
\newtheorem{lemma}{Lemma}
\newtheorem{cor}{Corollary}
\newtheorem{definition}{Definition}
\newtheorem{theorem}{Theorem}
\newtheorem{prop}{Proposition}
\newtheorem{conj}{Conjecture}
\maketitle
\begin{abstract}
The quasi-particle structure of the higher spin XXZ model is studied.
We obtained a new description of crystals associated with
the level $k$ integrable highest weight $U_q(\widehat{sl_2})$
modules in terms of
the creation operators at $q=0$ (the crystaline spinon basis).
The fermionic character formulas and the Yangian structure
of those integrable modules naturally follow from this
description.
We have also derived the conjectural formulas for
the multi quasi-particle states at $q=0$.
\end{abstract}
\par

\noindent
{\bf Introduction}
\par
\noindent
In this paper, we consider the integrable XXZ spin chain
with spin $k/2$ of ${sl_2}$.
The space of states is the infinit tensor product
(the space of local fields)
$$
{\cal W}=\cdots \otimes {\bf C}^{k+1} \otimes {\bf C}^{k+1} \otimes
{\bf C}^{k+1} \otimes {\bf C}^{k+1} \otimes \cdots.
$$
In remarkable papers \cite{FT,T}, using the Bethe Ansatz,
Fadeev and Takhatajan discovered that the one-particle excitation
in the anti-ferromagnetic regime of XXX chain
is always (a kink of) spin $1/2$.
According to this picture,
one can expect another description of space of states\cite{FT,R}
such as (the space of asymptotic particles)
$$
{\cal F}=\sum_{n=0}^{\infty}
\Big[ \sum_{p \in {\rm path}}
{\bf C}((z_1,\cdots,z_n)) \otimes
  [\otimes^{n} {\bf C}^2]_p \Big]^{\rm Symm},
$$
where Symm is the symmetrization with respect to the S-matrix.
On the other hand, stimulated by the deep results by Smirnov\cite{S},
the third description of the space of states in terms of the
representations of $U_q(\widehat {sl_2})$
(the space of non-local symmetries) was proposed\cite{DFJMN,IIJMNT}
$$
{\cal H}=\sum_{i,j=0}^{k} V(\la_i) \hat{\otimes} V(\la_j)^{\ast}.
$$
Here $V(\la_i)$ is the integrable highest weight representation of
$U_q(\widehat {sl_2})$ and $\hat{\otimes}$ means some extended tensor
product.
The particle picture is recovered using the $q$-vertex operators
acting among them.

In \cite{BLS1,BLS2}, a new description of the Hilbert space of the (chiral)
WZW conformal field theory was obtained.
The spin $1/2$ vertex operators were identified with the particles
(spinons) and the basis of the integrable ${\widehat {sl_2}}$ modules are
determined in terms of the spinons (spinon basis).
Their picture is clear, but technically it is quite hard
because of very complicated algebra of spinons.

The aim of this paper is to establish the remarkable equivalence
$$
{\cal H}={\cal F},
$$
rigorously, focusing on its combinatrial aspects, that is at the
level of crystals.
In a sense this can be considered as the crystalization
of the $q$-deformation of \cite{BLS1,BLS2}

One of (and the most remarkable) advantages to consider $q$-analogue
is that one can go into $q \rightarrow 0$ limit where everything
becomes clear and transparent (the "crystal" theory).
It is expected that in the limit $q \rightarrow 0$
the resulting algebraic structures for
"crystaline spinon" will be much simpler than that of $q=1$.

Here we must recall that there are two kinds of $q$-deformation
of the vertex operators, called type I and type II in the terminology
of \cite{DFJMN}.
They almost have the same properties as far as we consider them separately.
However, since it is crucial, in the study of spin chains, to
consider them simultaneously, they inevitably reveal different
faces.

While the type I $q$-vertex operators
have a well defined $q=0$ limit and play the central role
for the equivalence ${\cal H}={\cal W}$,
the existence of $q \rightarrow 0$ limits of the creation operators
(those are represented as type II $q$-VO) is not clear.
Indeed it is known that type II $q$-VO
produces poles at $q=0$, hence the $q \rightarrow 0$ limit does
{\it not} exist in a naive sence.

Nevertheless, it is conjectured \cite{DFJMN} that
those poles summed up to a meromorphic function in spectral parameters
and the $q \rightarrow 0$ limits of the creation operators are well-defined,
if they act on the {\it true vacuum}, the ground state of the model.
In fact what we need here is this type II $q$-vertex operators.

One way to avoid this subtle problem
is to construct the creation operators and
their action on the path basis directly at $q=0$,
rather than taking the limit $q \rightarrow 0$.
For the level $k=1$ case, this kind of description of $q=0$
creation operators was considered previously in \cite{DFJMN},
and we will generalize it to higher levels $k \geq 2$.
Though there still remains a deep gap between $q=0$ and $q \neq 0$
descriptions,
thus obtained statements of combinatrial nature (such as
character formulas) are indepedent of $q$ and available to generic $q$
including $1$.
This is reminiscent of "Combinatrial Bethe Anzats" due to \cite{KR}.

The paper is organized as follows.
Section 1 is a quick review of the path realization that shows
the eqivalence ${\cal H}={\cal W}$.
Then in section 2, we define and study the algebra of the creation
operators at $q=0$, and give a precise combinatrial description
of the space ${\cal F}$ including subtleties on the statistics.
In section 3, we formulate and prove our main theorem
(Theorem \ref{base}) that gives bijection (isomorphism of crystals)
between space of fields ${\cal W}$ and space of particles ${\cal F}$.
In section 4 and 5, we discuss about the Yangian structure and
spinon (or fermionic) character formulas.
These kinds of structures, that are conjectured by recent
TBA analysis (see for example\cite{KMM,KNS,K}),
will be obtained as corollaries of the results in section 3.
In section 6 the conjectural formulas for the $q=0$ limit of the
quasi-particles of the spin $k/2$ XXZ model are given
in terms of path basis. This is also an easy consequence
of the commutation relations of creation operators at $q=0$.
Section 7 is devoted to comments and discussions.

\section{Review of the path realization}
\par
Here we review a path realization of crystals.
Unless otherwise stated we follow the notations
of $\S2$ of \cite{IIJMNT}. Other notations and
fundamental properties of crystals which we use in this paper
we refer \cite{KKMMNN1,N}.
We denote by $B^{(k)}=\{b^{(k)}_l|0\leq l\leq k\}$
the associated crystal to
the crystal base of the $k+1$ dimensional
$U_q({\widehat{sl_2}})$ module $V^{(k)}$.
The actions of $\et_i$, $\ft_i$ $(i=0,1)$ on $B^{(k)}$
are specified by
\begin{eqnarray}
&&
\ft_1b^{(k)}_l=\et_0b^{(k)}_l=b^{(k)}_{l+1},
\quad 0\leq l\leq k-1,
\nonumber
\\
&&
\et_1b^{(k)}_l=\ft_0b^{(k)}_l=b^{(k)}_{l-1},
\quad 1\leq l\leq k,
\nonumber
\end{eqnarray}
and $\tilde{x}_ib^{(k)}_l=0$ (otherwise) for $x=e,f$, $i=0,1$.
We call a crystal defined from $U_q({\widehat{sl_2}})$
${\widehat{sl_2}}$ crystal.
A ${\widehat{sl_2}}$ crystal is called
$({\widehat{sl_2}})_i$ crystal $(i=0,1)$
if we forget the actions of $\et_{1-i}$, $\ft_{1-i}$,
$\{h_i\}$ being the simple coroots.
Let $P$ and $P^\ast$ denote the weight and the dual weight
lattice of ${\widehat{sl_2}}$,
$P=\bz\La_0\oplus\bz\La_1\oplus\bz\delta$,
$P^\ast=\bz h_0\oplus\bz h_1\oplus\bz d$,
$\langle h_i,\La_j \rangle=\delta_{ij}$,
$\langle d,\delta \rangle=1$,
$\langle h_i,\delta \rangle=\langle d,\La_j \rangle=0$.

Let us set $\cp^{(m,m^{'})}=B(\la_m)\ot B(\la_{m^{'}})$,
which is the crystal associated to $V(\la_m) \ot V(\la_{m^{'}})^{\ast}$.
The ground state path $\bp_m$ is defined by
\begin{eqnarray}
&&
\bp_m(l)=m+(k-2m)\ep(l)
\qquad
l\in\bz,
\nonumber
\end{eqnarray}
where $\ep(l)=0(l:even)=1(l:odd)$. Define the $(m,m^{'})$
ground state path by
$$
\bp_{m,m^{'}}(l)=
\left\{
\begin{array}{ll}
\bp_m(l) & \quad l\geq1,\\
\bp_{m^{'}}(l) & \quad l<0.
\end{array}
\right.
$$
Then

\begin{prop}
There is a bijection
\begin{eqnarray}
&&
\cp^{(m,m^{'})}=
\{
p=(p(l))_{l\in\bz} \ | \
p(l)\in\{0,\cdots,k\},
p(l)=\bp_{m,m^{'}}(l) \ (|l|>>0)
\}.
\nonumber
\end{eqnarray}
\end{prop}
We call an element in the right hand side of this
$(m,m^{'})$ path. From now on we identify $\cp^{(m,m^{'})}$
with the set of $(m,m^{'})$ paths.

A $(m,m^{'})$ path can be described in terms of domain walls.
For $p\in\cp^{(m,m^{'})}$ we can associate the uniquely
determined data
\begin{eqnarray}
&&
n\in\bz_{\geq0},
\quad
(l_n,\cdots,l_1)\in\bz^n,
\nonumber
\\
&&
\hbox{and }
(m_n,\cdots,m_0)\in\{0,\cdots,k\}^{n+1}
\nonumber
\end{eqnarray}
such that
\begin{eqnarray}
&&
l_n\geq\cdots\geq l_1,
\nonumber
\\
&&
m_n=m,\quad m_0=m^{'},
\nonumber
\\
&&
|m_j-m_{j-1}|=1\quad\hbox{ for any $j$},
\nonumber
\\
&&
\hbox{if $l_{s+s^{'}}>l_{s+s^{'}-1}=\cdots=l_{s}>l_{s-1}$,
then $|m_{s+s^{'}-1}-m_s|=s^{'}$},
\nonumber
\\
&&
p(l)=\bp_{m_r}(l)\qquad l_{r+1}\geq l\geq l_r+1.
\label{cpath}
\end{eqnarray}
If $l_r=l_{r+1}$, we understand that there are no corresponding
condition of the form (\ref{cpath}).
Conversely these data uniquely determines the
$(m,m^{'})$ path. For a fixed sequence
$(m_n,\cdots,m_0)$ we denote this path by
$[[l_n,\cdots,l_1]]$.

Let us describe the actions of $\et_i$ and $\ft_i$ on
$[[l_n,\cdots,l_1]]$.
Set
\begin{eqnarray}
&&
s_j={\rm sign}(p(l_j)+p(l_{j}+1)-k) \quad 1\leq j\leq n
\label{wall1}
\end{eqnarray}
and $r_j={1\over2}(1+s_j)$, where
${\rm sign}(j)=1(j>0)=-1(j<0)$. Note that
$p(l_j)+p(l_{j}+1)-k\neq 0$.
Let us associate the element $b$
of $B^{(1)\ot n}$ with the path $p$ by
\begin{eqnarray}
&&
b=b^{(1)}_{r_n}\ot\cdots\ot b^{(1)}_{r_1}.
\label{wall2}
\end{eqnarray}
Suppose that
\begin{eqnarray}
&&
\tilde{x}_ib=
b^{(1)}_{r_n}
\ot\cdots\ot
\tilde{x}_ib^{(1)}_{r_j}
\ot\cdots\ot
b^{(1)}_{r_1}.
\label{wall3}
\end{eqnarray}
Then
\begin{equation}
\tilde{x}_i
{[[} l_n,\cdots,l_1 {]]}=
\left \{
\begin{array}{cc}
0, & \quad {\rm if} \ \tilde{x}_ib=0, \\
{[[} l_n^{'},\cdots,l_1^{'} {]]}, & \quad {\rm if} \ \tilde{x}_ib\neq0,
\end{array}
\right.\label{wall4}
\end{equation}
where $l_s^{'}=l_s$ $(s\neq j)$ and
\begin{equation}
l_j^{'}=
\left \{
\begin{array}{ll}
l_j+1, & \quad {\rm for} \ x=f, \\
l_j-1, & \quad {\rm for} \ x=e.
\end{array}
\right.\label{wall5}
\end{equation}

The weight of a $(m,m^{'})$ path $p$ is given by
\begin{eqnarray}
&&
wt(p)=
(m-m^{'}+2s(p))(\La_1-\La_0)-h(p)\delta,
\nonumber
\\
&&
s(p)=\sum_{l\in \bz}(\bp_{m,m^{'}}(l)-p(l)),
\nonumber
\\
&&
h(p)=
\sum_{l\in\bz}
l\Big(
\th(p(l+1),p(l))-
\th(\bp_{m,m^{'}}(l+1)-\bp_{m,m^{'}}(l))
\Big),
\nonumber
\end{eqnarray}
where the function $\th(j,j^{'})$ is defined by
$$
\th(j,j^{'})=
\left \{
\begin{array}{ll}
-j^{'} & \quad {\rm if} \ j+j^{'}\leq k, \\
j-k    & \quad {\rm if} \ j+j^{'}> k.
\end{array}
\right.
$$

\section{Creation algebra at $q=0$}
\par
\begin{definition}
The algebra $\ca$ is generated by
$\{\v^{\ast p}_j \ | \ j\in\bz,\,p\in\{0,1\}\}\cup\{1\}$
over $\bz$ subject to the
following defining relations:
\begin{eqnarray}
&
\v_{j_1}^{\ast p_1}\v_{j_2}^{\ast p_2}+
\v_{j_2}^{\ast p_1}\v_{j_1}^{\ast p_2}=0,
&
\hbox{ $j_1=j_2$ {\rm mod.}$2$ {\rm and} $(p_1,p_2)=(1,0)$},
\nonumber
\\
&
\v_{j_1}^{\ast p_1}\v_{j_2}^{\ast p_2}+
\v_{j_2+1}^{\ast p_1}\v_{j_1-1}^{\ast p_2}=0,
&
\hbox{ $j_1\neq j_2$ {\rm mod.}$2$ {\rm and} $(p_1,p_2)=(1,0)$},
\nonumber
\\
&
\v_{j_1}^{\ast p_1}\v_{j_2}^{\ast p_2}+
\v_{j_2-2}^{\ast p_1}\v_{j_1+2}^{\ast p_2}=0,
&
\hbox{ $j_1=j_2$ {\rm mod.}$2$ {\rm and} $(p_1,p_2)\neq(1,0)$},
\nonumber
\\
&
\v_{j_1}^{\ast p_1}\v_{j_2}^{\ast p_2}+
\v_{j_2-1}^{\ast p_1}\v_{j_1+1}^{\ast p_2}=0,
&
\hbox{ $j_1\neq j_2$ {\rm mod.}$2$ {\rm and} $(p_1,p_2)\neq(1,0)$},
\nonumber
\\
&
\v_{j}^{\ast p}1=1\v_{j}^{\ast p}=\v_{j}^{\ast p}.
\qquad\quad\,
&
\nonumber
\end{eqnarray}
\end{definition}

As a special case of the above defining equations we have
\begin{eqnarray}
&&
\v_j^{\ast p_1}\v_j^{\ast p_2}=
\v_j^{\ast p_1}\v_{j-1}^{\ast p_2}=0,
\hbox{ if $(p_1,p_2)=(1,0)$},
\nonumber
\\
&&
\v_j^{\ast p_1}\v_{j+1}^{\ast p_2}=
\v_j^{\ast p_1}\v_{j+2}^{\ast p_2}=0,
\hbox{ if $(p_1,p_2)\neq(1,0)$}.
\nonumber
\end{eqnarray}

We call $\ca$ the crystaline creation algebra.
The derivation of the above commutation relations
is explained in appendix A.
The algebra $\ca$ is naturally graded by
\begin{eqnarray}
&&
\ca=\oplus_{n=0}^\infty \ca_n
\nonumber
\\
&&
\ca_n=\sum_{(p_n,\cdots,p_1),(j_n,\cdots,j_1)}
\bz\v^{\ast p_n}_{j_n}\cdots\v^{\ast p_1}_{j_1},
\quad
\ca_0=\bz.
\end{eqnarray}

Let us introduce the functions $H$, $K$, $\ep$ by
\begin{eqnarray}
&&
H(0,0)=H(0,1)=H(1,1)=0,\quad H(1,0)=1
\nonumber
\\
&& K(p_1,p_2)=1-H(p_1,p_2),\quad
\ep(j)={1-(-1)^j\over2}.
\nonumber
\end{eqnarray}
Then the commutation relations can be written compactly as
\begin{eqnarray}
&&
\v^{\ast p_2}_{j_2}\v^{\ast p_1}_{j_1}+
\v^{\ast p_2}_{j_1-2K(p_2,p_1)+\ep(j_1+j_2)}
\v^{\ast p_1}_{j_2+2K(p_2,p_1)-\ep(j_1+j_2)}=0.
\nonumber
\end{eqnarray}
Let us set
\begin{eqnarray}
&&
B=\cup_{n=1}^\infty
\{\v_{j_n}^{\ast p_n}\cdots\v_{j_1}^{\ast p_1}|
j_l\in\bz,\,p_l\in\{0,1\}\},
\nonumber
\\
&&
B(p_n,\cdots,p_1)=
\{\v_{j_n}^{\ast p_n}\cdots\v_{j_1}^{\ast p_1}|
\hbox{ $(j_n,\cdots,j_1)$ satisfies the condition (\ref{nor})}\},
\nonumber
\\
&&
B_{\geq0}(p_n,\cdots,p_1)=
\{\v_{j_n}^{\ast p_n}\cdots\v_{j_1}^{\ast p_1}\in B(p_n,\cdots,p_1)|
j_1\geq0\}.
\nonumber
\end{eqnarray}
If $n=0$, we define $(p_n, \cdots, p_1)=\phi$ and $B(\phi)=\{ 1 \}$.
The condition is
\begin{eqnarray}
&&
j_n-2I_n(p_n,\cdots,p_1)\geq\cdots\geq
j_2-2I_2(p_2,p_1)\geq j_1,
\label{nor}
\\
&&
I_l(p_l,\cdots,p_1)=\sum_{s=1}^{l-1}H(p_{s+1},p_s).
\nonumber
\end{eqnarray}

\begin{theorem}\label{linear}
$B(p_n,\cdots,p_1)$ is a $\bz$ linear base of $\ca_n$.
\end{theorem}

We prove the theorem in a more general setting.
Let $J$ be a set.
Let us consider the associative algebra over $\bz$
with unit generated by $\{\psi_A(n)|A\in J,n\in\bz\}$
whose defining relations are
$$
\psi_A(n) \psi_B(m)+\psi_A(m+s_{AB}) \psi_B(n-s_{AB})=0,
$$
where $s_{AB}$ is some fiexd integer.
In what follows, we fix a sequence $(A_1,\cdots,A_n)$
and put $s_{i,i+1}=s_{A_i,A_{i+1}}$. In general let us define
$$
s_{i,j}=
\left \{
\begin{array}{cc}
s_{i,i+1}+s_{i+1,i+2}+\cdots+s_{j-1,j}, & \quad (i<j), \\
0, & \quad (i=j), \\
-s_{j,i}, & \quad (i>j).
\end{array}
\right.
$$

\begin{lemma}\label{permu}
For any permutation $\sigma$ of $(1,2,\cdots,n)$,
$$
\psi_{A_1}(m_1)\cdots\psi_{A_n}(m_n)=
{\rm sgn}(\sigma) \psi_{A_1}(m_{1\sigma(1)})\cdots\psi_{A_n}(m_{n\sigma(n)}),
$$
where
$$
m_{i,j}=m_{j}+s_{i,j}.
$$
\end{lemma}
\par
\noindent
Proof
\par
\noindent
By the inducion on the length of $\sigma$, the lemma is easily
proved. $\Box$

\begin{definition}
In a product
$$
\Psi=\psi_{A_1}(m_1)\cdots\psi_{A_n}(m_n),
$$
the index pair $(m_i,m_j)$ $(i<j)$ is said to be normal (zero, abnormal) iff
$$
m_i > m_j+s_{i,j} \quad (m_i = m_j+s_{i,j}, \quad m_i < m_j+s_{i,j}).
$$
Furthermore, $\Psi$ is normal iff all pairs are normal.
\end{definition}

\begin{lemma}\label{nnormal}
The notion of normal (zero, abnormal) of two indices $(m_i,m_j)$
is independent of the position of them as far as their
order is preserved. If the order is exchanged, then the normal pair
is transformed into abnormal one and zero pair to zero pair.
\end{lemma}
\par
\noindent
Proof
\par
\noindent
Under any permutation $\sigma$ such that $\sigma^{-1}(i)=i'$
and $\sigma^{-1}(j)=j'$, one has
$$
m_{i',i}=m_i+s_{i',i}, \quad m_{j',j}=m_j+s_{j',j}.
$$
Then, for the new pair $(m_{i',i},m_{j',j})$ at position $(i',j')$, one has
$$
m_{i',i}-m_{j',j}-s_{i',j'}=m_i-m_j+s_{i',i}-s_{i'j'}-s_{j',j}=m_i-m_j-s_{i,j},
$$
and
$$
m_{j',j}-m_{i',i}-s_{j',i'}=-(m_i-m_j-s_{i,j}).
$$
Hence the normal (zero, abnormal)
pair is transformed into
normal (zero, abnormal) for $i'<j'$, or abnormal (zero, normal)
for $i'>j'$.
$\Box$

\begin{cor}\label{normal}
A product of the form
$$
\Psi=\psi_{A_1}(m_1)\cdots\psi_{A_n}(m_n),
$$
is $0$ iff there exist (al least) a pair $(m_i,m_j)$ such that
$$
m_i = m_j+s_{i,j}.
$$
Othewise, it can be transformed into normal form.
\end{cor}

\begin{cor}\label{nbase}
The set of normal forms
$\{\psi_{A_1}(m_1)\cdots\psi_{A_n}(m_n)\}$
forms  a $\bz$ linear base of the algebra.
\end{cor}
\par
\noindent
Proof
\par
\noindent
The corollary is proved by the standard argument
of constructing a representation of the algebra
using Corollary \ref{normal}.
$\Box$
\vskip3truemm
\par
\noindent
Proof of Theorem \ref{linear}
\par
\noindent
Take $J=\{0,1\}^2$ and
$$
s_{(p_1,i_1)(p_2,i_2)}=-K(p_1,p_2)+H(i_2,i_1).
$$
Then the above algebra is isomorphic to $\ca$ by
$\v^{\ast p}_{2n+i}=\psi_{(p,i)}(n)$, where $i=0,1$.
In this case the normality condition is exactly the condition (\ref{nor}).
Hence Theorem \ref{linear} follows from Corollary \ref{nbase}.
$\Box$

\begin{definition}
Let us define the weight of a nonzero element of $B$ by
\begin{eqnarray}
wt(\v^{\ast p_n}_{j_n}\cdots\v^{\ast p_1}_{j_1})&=&
\sum_{l=1}^nwt(\v^{\ast p_l}_{j_l})
\nonumber
\\
wt(\v^{\ast p}_{j})&=&
-[{j\over2}]\delta+(1-2\ep(j))(\La_1-\La_0),
\nonumber
\end{eqnarray}
where $[j]$ is the Gauss symbol. We set $wt1=0$.
\end{definition}

Let us introduce the actions of $\et_i$, $\ft_i$ $(i=0,1)$
on the set $B(p)=\{\v_j^{\ast p}|j\in\bz\}$ $(p=0,1)$ by
\begin{eqnarray}
&&
\ft_1\v_{2j}^{\ast p}=\v_{2j+1}^{\ast p},
\quad
\ft_0\v_{2j-1}^{\ast p}=\v_{2j}^{\ast p},
\nonumber
\\
&&
\et_1\v_{2j+1}^{\ast p}=\v_{2j}^{\ast p},
\quad
\et_0\v_{2j}^{\ast p}=\v_{2j-1}^{\ast p},
\nonumber
\\
&&
\tilde{x}_i\v_{j}^{\ast p}=0\qquad
\hbox{otherwise},
\nonumber
\end{eqnarray}
where $x=e,f$.
By these actions and the weight of $\v_{j}^{\ast p}$,
$B(p)$ is an  affine crystal\cite{KKMMNN1,N}
isomorphic to ${\rm Aff}(B^{(1)})$.
In general

\begin{theorem}\label{crystal}
Let $n\geq1$ and $(p_n,\cdots,p_1)\in\{0,1\}^n$.
There is a unique ${\widehat{sl_2}}$ crystal structure
in $B(p_n,\cdots,p_1)$ such that
the natural map
\begin{eqnarray}{\rm Aff}(B^{(1)})^{\ot n}
&\longrightarrow&
\Big(B(p_n,\cdots,p_1)\cup-B(p_n,\cdots,p_1)\cup\{0\}\Big)\Big/\pm
\nonumber
\\
\v^{\ast p_n}_{j_n}\ot\cdots\ot\v^{\ast p_1}_{j_1}
&\longmapsto&
\v^{\ast p_n}_{j_n}\cdots\v^{\ast p_1}_{j_1}
\nonumber
\end{eqnarray}
commutes with the actions of $\et_i$, $\ft_i$ $(i=0,1)$.
\end{theorem}
We consider $B(0)=\{1\}$ to be the trivial crystal,
$\tilde{x}_i1=0$ for $x=e,f$ and $i=0,1$.
\vskip5truemm

\par
\noindent
Proof of Theorem \ref{crystal}
\par
\noindent
Take any $(p_2,p_1)\in \{0,1\}^2$ and fix it.
Let us set, as a subset of the tensor algebra
generated by $B(p)$ $(p=0,1)$,
\begin{eqnarray}
&&
{\cal I}=
\{
\v^{\ast p_2}_{j_2}\ot\v^{\ast p_1}_{j_1}
+
\v^{\ast p_2}_{j_1-2K(p_2,p_1)+\ep(j_1+j_2)}
\ot\v^{\ast p_1}_{j_2+2K(p_2,p_1)-\ep(j_1+j_2)}|
j_1,j_2\in\bz
\}
\nonumber
\end{eqnarray}
Then it is sufficient to prove that
\begin{eqnarray}
&&
\tilde{x}_i{\cal I}\subset{\cal I}\sqcup\{0\}
\nonumber
\end{eqnarray}
for $x=e,f$, $i=0,1$.
Here in general for the elements $b_1,b_2$ of a crystal
we understand
$\tilde{x}_i(b_1+b_2)=\tilde{x}_ib_1+\tilde{x}_ib_2$.
By direct calculations this
property is easily proved.
$\Box$

\section{Crystaline spinon base}
\par

\begin{definition}
Let us call $(p_n,\cdots,p_1)\in\{0,1\}^{n}$ a
level $k$ restricted path from $0$ to
$l$ of length $n$ if
\begin{eqnarray}
&&
(-1)^{p_1}+\cdots+(-1)^{p_s}
\in\{0,\cdots,k\}
\hbox{ for $1\leq s\leq n$}.
\nonumber
\\
&&(-1)^{p_1}+\cdots+(-1)^{p_n}=l.
\nonumber
\end{eqnarray}
We denote by $\pek_{res,n}(0,l)$ the set of restricted paths
from $0$ to $l$ of length $n$ and set
$$
\pek_{res,n}=\sqcup_{l=0}^k\pek_{res,n}(0,l).
$$
\end{definition}
\par
\noindent
We understand that $\pek_{res,0}=\{ \phi \}$.

The following theorem provides a new parametrization
of the crystal base of the integrable highest or lowest weight
$U_q({\widehat{sl_2}})$ modules.

\begin{theorem}\label{base}
There is an isomorphism of affine crystals
\begin{eqnarray}
&&
\sqcup_{n=0}^\infty\sqcup_{(p_n,\cdots,p_1)\in \pek_{res,n}(0,l)}
B(p_n,\cdots,p_1)
\simeq
B(\la_l)\ot B(\la_0)^\ast
\nonumber
\end{eqnarray}
given by
\begin{eqnarray}
1&\longmapsto&[[ \ ]]=b_{\la_0}\ot b_{-\la_0},
\nonumber
\\
\v^{\ast p_n}_{j_n}\cdots\v^{\ast p_1}_{j_1}
&\longmapsto&
[[j_n-p_n,\cdots,j_1-p_1]].
\nonumber
\end{eqnarray}
\end{theorem}

\begin{cor}\label{hbase}
The map in Theorem \ref{base} induces the isomorphism
of $({\widehat{sl_2}})_1$ crystals with affine weights:
\begin{eqnarray}
&&
\sqcup_{n=0}^\infty\sqcup_{(p_n,\cdots,p_1)\in \pek_{res,n}(0,l)}
B_{\geq0}(p_n,\cdots,p_1)
\simeq
B(\la_l),
\nonumber
\end{eqnarray}
where we make the identification:
\begin{eqnarray}
B(\la_l)&\simeq& B(\la_l)\ot b_{-\la_0}
\nonumber
\\
b&\longrightarrow& b\ot b_{-\la_0}.
\nonumber
\end{eqnarray}
Here $b_{-\la_0}$ is the lowest weight element in $B(\la_0)^\ast$.
\end{cor}
\par
\noindent
Note that there are no naturally defined ${\widehat{sl_2}}$ crystal
structure on $B_{\geq0}(p_n,\cdots,p_1)$.

The map in Theorem \ref{base} determines a representation
of ${\cal A}$ as in the proof of Corollary \ref{nbase}.
The action of $\v^{\ast p}_j$ on $\oplus_{b}\bz b$ is described
in the following manner, where $b$ runs over all the elements
in $\sqcup_{l=0}^k B(\la_l)\ot B(\la_0)^\ast$.
Let $b=[[j_{n-1}-p_{n-1},\cdots,j_1-p_1]]$ be as in Theorem \ref{base}
and
$\v=\v^{\ast p}_j\v^{\ast p_{n-1}}_{j_{n-1}}\cdots\v^{\ast p_1}_{j_1}$.
If $\v\neq0$ then, by Lemma \ref{permu} and Lemma \ref{nnormal},
there exist unique $\mu\in\{\pm1\}$ and
normal sequence $(j_n^{'},\cdots,j_1^{'})$ such that
\begin{eqnarray}
&&
\v=
\mu \v^{\ast p}_{j_n^{'}}\v^{\ast p_{n-1}}_{j_{n-1}^{'}}
\cdots\v^{\ast p_1}_{j_1^{'}}.
\nonumber
\end{eqnarray}
We define $\v^{\ast p}_jb=0$ if $(p,p_{n-1},\cdots,p_1)$
is not a restricted path.
Suppose that $(p,p_{n-1},\cdots,p_1)\in\pek_{res,n}$.
Then
$$
\v^{\ast p}_jb=
\left \{
\begin{array}{cc}
0, & \quad \hbox{if $\v=0$}, \\
\mu [[j_n^{'}-p,j_{n-1}^{'}-p_{n-1},\cdots,j_1^{'}-p_1]],
& \quad \hbox{if $\v\neq0$}.
\end{array}
\right.
$$
In particular for the element in $B(p_n,\cdots,p_1)$ we have
\begin{eqnarray}
&&
\v^{\ast p_n}_{j_n}\cdots\v^{\ast p_1}_{j_1}[[\,]]
=
[[j_n-p_n,\cdots,j_1-p_1]].
\nonumber
\end{eqnarray}
This representation is considered as a Fock type representation
of the crystaline creation algebra, the vacuum being $[[\,]]$.

We call the basis given in Theorem \ref{base}
and Corollary \ref{hbase} the crystaline spinon basis.
\vskip2truemm

\noindent
Proof of Theorem \ref{base}
\par
\noindent
The bijectivity of the map is obvious by the description
of the paths in terms of domain walls and the normality condition
(\ref{nor}).
So what we should prove is the following two statements.
\begin{description}
\item[(i)]
The map commutes with the actions of $\et_i$, $\ft_i$ $(i=0,1)$.
\item[(ii)]
The map preserves weights.
\end{description}

Let us prove (i) first.

\begin{lemma}\label{domain}
Let $p=(p(n))_{n\in\bz}=[[j_n-p_n,\cdots,j_1]]$ be
an element of $\sqcup_{l=0}^k B(\la_l)\ot B(\la_0)^\ast$.
If
$$
j_{r+l}-p_{r+l}>
j_{r+l-1}-p_{r+l-1}=\cdots=j_r-p_r>j_{r-1}-p_{r-1}
$$
then
\begin{eqnarray}
&&
p(j_r-p_r)+p(j_r-p_r+1)=k-l(-1)^{j_r}.
\nonumber
\end{eqnarray}
\end{lemma}
Note that the assumption of the lemma implies
$p_{r+l-1}=\cdots=p_r$ and $j_{r+l-1}=\cdots=j_r$.
By the rule (\ref{wall1}) - (\ref{wall3}),
the above domain wall corresponds to
$b^{\ot(1)}_{\ep(j_r)} \in B^{(1) \otimes l}$.
\par
\noindent
Proof
\par
\noindent
The statement is easily proved by direct calculations.
$\Box$

Let us consider the classical crystal morphism
\begin{eqnarray}
cl^n: {\rm Aff}(B^{(1)\ot n})&\longrightarrow& B^{(1)\ot n},
\nonumber
\\
\v^{\ast p_n}_{j_n}\ot\cdots\ot\v^{\ast p_1}_{j_1}
&\mapsto&
b^{(1)}_{\ep(j_n)}\ot\cdots\ot b^{(1)}_{\ep(j_1)}.
\nonumber
\end{eqnarray}
Since $cl^n$ commutes with the actions of $\et_i$ and $\ft_i$
\begin{eqnarray}
&&
\tilde{x}_i(\v^{\ast p_n}_{j_n}\ot\cdots\ot\v^{\ast p_1}_{j_1})
=
\v^{\ast p_n}_{j_n}
\ot\cdots\ot
\tilde{x}_i\v^{\ast p_l}_{j_l}
\ot\cdots\ot
\v^{\ast p_1}_{j_1}
\nonumber
\end{eqnarray}
is equivalent to
\begin{eqnarray}
&&
\tilde{x}_i(b^{(1)}_{\ep(j_n)}\ot\cdots\ot b^{(1)}_{\ep(j_1)})
=
b^{(1)}_{\ep(j_n)}
\ot\cdots\ot
\tilde{x}_i b^{(1)}_{\ep(j_l)}
\ot\cdots\ot
b^{(1)}_{\ep(j_1)}.
\nonumber
\end{eqnarray}
Hence (i) follows from the definition of the actions
of $\et_i$, $\ft_i$ on $B(p)$ and (\ref{wall1})-(\ref{wall5}).
$\Box$

Next let us prove (ii).

\begin{lemma}\label{separ}
Let $p\in\sqcup_{m,m^{'}}\cp^{(m,m^{'})}$
be a path with $n$ domain walls counting multiple ones.
Then there exists $P$ which is a composition
of $\et_i$ and $\ft_i$ $(i=0,1)$ such that
\begin{eqnarray}
&&
Pp=[[l_n,\cdots,l_1]],
\quad
l_n>\cdots>l_1\geq0.
\nonumber
\end{eqnarray}
\end{lemma}
\par
\noindent
Proof
\par
\noindent
Since $B^{(1)}$ is a perfect crystal of level $1$,
$B^{(1)\ot n}$ is connected. Hence there exists a composition
$P_1$ of $\et_i$, $\ft_i$ $(i=0,1)$ such that
\begin{eqnarray}
&&
cl^n(P_1p)=b^{(1)\ot n}_0.
\nonumber
\end{eqnarray}
If we denote $P_1p=[[l_n,\cdots,l_1]]$ then
\begin{eqnarray}
&&
\ft_0^{n}\ft_1^nP_1p=[[l_n+2,\cdots,l_1+2]].
\nonumber
\end{eqnarray}
Hence $p$ is connected to $[[r_n,\cdots,r_1]]$
such that
\begin{eqnarray}
&&
r_1\geq0,
\quad
cl^n[[r_n,\cdots,r_1]]=b^{(1)\ot n}_0.
\nonumber
\end{eqnarray}
So let us assume that $p$ is of this form from the beginning.
Then
\begin{eqnarray}
(\ft_0\ft_1^2)(\ft_0^3\ft_1^4)\cdots(\ft_0^{n-2}\ft_1^{n-1})p
&=&[[r_n+n-1,\cdots,r_2+1,r_1]]
\quad
\hbox{$n$: odd},
\nonumber
\\
\ft_1(\ft_0^2\ft_1^3)\cdots(\ft_0^{n-2}\ft_1^{n-1})p
&=&[[r_n+n-1,\cdots,r_2+1,r_1]]
\quad
\hbox{$n$: even}
\nonumber
\end{eqnarray}
which proves the lemma.
$\Box$
\vskip2truemm
\noindent

We prove the weight preservation
by induction on the number of domain walls.
By Lemma \ref{separ} and the statement (i) above it is sufficient
to prove the weight preservation for such a path
$[[l_n,\cdots,l_1]]$
that
\begin{eqnarray}
&&
l_n>\cdots>l_1\geq0.
\nonumber
\end{eqnarray}
\par
\noindent
0)
For the vaccum (=zero domain wall) state, the statement is obvious
since $wt([[ \ ]])=wt(1)=0$.
Now, we will prove that
$$
wt(p_2)-wt(p_1)=wt({\cal O}),
$$
for any $(b,c)$- and $(a,c)$-path, $p_1$ and $p_2$ such as
$$
p_1(l)=\left\{ \begin{array}{ll}
                \bp_b(l), & n < l, \\
                q(l),     & l \leq n, \\
               \end{array} \right.
\quad {\rm and} \quad
p_2(l)=\left\{ \begin{array}{ll}
                \bp_a(l), & m < l, \\
                \bp_b(l), & n < l \leq m, \\
                q(l),     & l \leq n, \\
               \end{array} \right.
$$
and the corresponding creation operator ${\cal O}$
(see 3) below). Here we can assume $m>n\geq0$.

\par
\noindent
1)
First, calculate the $h$-functiom.
For $h(p_1)$, one decompose the summation into four parts
$h(p_1)=h_1^{(4)}+h_1^{(3)}+h_1^{(2)}+h_1^{(1)}$ as
$(n < l)$, $(n=l)$, $(n > l > 0)$ and $(0 >l)$, then
\begin{eqnarray}
&&h_1^{(4)}=
\sum_{n < l} l[\th(p_1(l+1),p_1(l))-\th(\bp_b(l+1),\bp_b(l))]=0,
\nonumber \\
&&h_1^{(3)}=
n[\th(\bp_b(n+1),q(n))-\th(\bp_b(n+1),\bp_b(n))],
\nonumber \\
\nonumber \\
&&h_1^{(2)}=
\sum_{n > l > 0} l[\th(q(l+1),q(l))-\th(\bp_b(l+1),\bp_b(l))],
\nonumber \\
&&h_1^{(1)}=
\sum_{0 > l} l[\th(q(l+1),q(l))-\th(\bp_c(l+1),\bp_c(l))].
\nonumber
\end{eqnarray}
Similarly, by decomposing to six parts, one has
$h(p_2)=h_2^{(6)}+\cdots +h_2^{(1)}$, where
\begin{eqnarray}
&&h_2^{(6)}=
\sum_{m < l} l[\th(p_2(l+1),p_2(l))-\th(\bp_a(l+1),\bp_a(l))]=0,
\nonumber \\
&&h_2^{(5)}=
m[\th(\bp_a(m+1),\bp_b(m))-\th(\bp_a(m+1),\bp_a(m))],
\nonumber \\
\nonumber \\
&&h_2^{(4)}=
\sum_{m > l > n} l[\th(\bp_b(l+1),\bp_b(l))-\th(\bp_a(l+1),\bp_a(l))],
\nonumber \\
&&h_2^{(3)}=
n[\th(\bp_b(n+1),q(n))-\th(\bp_a(n+1),\bp_a(n))],
\nonumber \\
\nonumber \\
&&h_2^{(2)}=
\sum_{n > l > 0} l[\th(q(l+1),q(l))-\th(\bp_a(l+1),\bp_a(l))],
\nonumber \\
&&h_2^{(1)}=
\sum_{0 > l} l[\th(q(l+1),q(l))-\th(\bp_c(l+1),\bp_c(l))].
\nonumber
\end{eqnarray}
Taking the difference of these two, one obtain
\begin{eqnarray}
h(p_2)-h(p_1)=
&&m[\th(\bp_a(m+1),\bp_b(m))-\th(\bp_a(m+1),\bp_a(m))]
\nonumber \\
&&+\sum_{m > l > 0} l[\th(\bp_b(l+1),\bp_b(l))-\th(\bp_a(l+1),\bp_a(l))].
\nonumber
\end{eqnarray}
The first line of the R.H.S. is evaluated as
$$
m[\th(\bp_a(m+1),\bp_b(m))-\th(\bp_a(m+1),\bp_a(m))]=
\left\{ \begin{array}{ll}
         m[\bp_a(m)-\bp_b(m)], & \bp_a(m) \geq \bp_b(m), \\
         0,                    & \bp_a(m) < \bp_b(m).
\end{array} \right.
$$
The second line is
$$
\sum_{n > l > 0} l[\th(\bp_b(l+1),\bp_b(l))-\th(\bp_a(l+1),\bp_a(l))]=
\left\{ \begin{array}{ll}
         -(a-b) m/2, & {\rm even} \ m, \\
          (a-b) (m-1)/2, & {\rm odd} \ m.
\end{array} \right.
$$
Hence
$$
h(p_2)-h(p_1)=\left\{
\begin{array}{lll}
(a-b) m/2, & {\rm even} \ m, & a>b,\\
-(a-b) m/2, & {\rm even} \ m, & a<b,\\
(a-b) (m-1)/2, & {\rm odd} \ m, & a>b,\\
-(a-b) (m+1)/2, & {\rm odd} \ m, & a<b.
\end{array} \right.
$$

\par
\noindent
2)
By similar and easier calculation, one can show
$$
s(p_2)-s(p_1)=\sum_{m \geq l > 0} [\bp_a(l)-\bp_b(l)]=
\left\{ \begin{array}{ll}
0, & {\rm even} \ m, \\
b-a, & {\rm odd} \ m.
\end{array} \right.
$$
Togehter with the results of 1), one obtain
$$
wt(p_2)-wt(p_1)=\left\{
\begin{array}{lll}
(a-b)(\La_1-\La_0)-(a-b) m/2 \ \delta, & {\rm even} \ m, & a>b,\\
(a-b)(\La_1-\La_0)+(a-b) m/2 \ \delta, & {\rm even} \ m, & a<b,\\
-(a-b)(\La_1-\La_0)-(a-b) (m-1)/2 \ \delta, & {\rm odd} \ m, & a>b,\\
-(a-b)(\La_1-\La_0)+(a-b) (m+1)/2 \ \delta, & {\rm odd} \ m, & a<b.
\end{array} \right.
$$

\par
\noindent
3)
On the other hand, in spinon basis, the path $p_2$ is obtained from $p_1$ by
applying the following (products of) operators
$$
{\cal O}=
\left\{
\begin{array}{ll}
(\v^{\ast 0}_{m})^{a-b}, &  a>b,\\
(\v^{\ast 1}_{m+1})^{b-a}, & a<b.
\end{array} \right.
$$
It is easy to see that the difference of $wt$, $wt(p_2)-wt(p_1)$
is given exactly by the weight of these operators, hence the theorem
is poved.$\Box$

\section{Yangian like structure}
\par
In this section we give a decomposition of
the crystal of integrable irreducible
highest weight $U_q({\widehat{sl_2}})$ modules as
$({\widehat{sl_2}})_1$ crystals
which has a natural description in terms of
crystaline spinon basis.
Those $({\widehat{sl_2}})_1$ crystals can be considered as describing the
Yangian module structure in the $q\ra1$ limit.
In fact one half of the character formula
(\ref{character}) corresponds to
this structure which is known to describe the Yangian contribution \cite{BLS2}.

\begin{definition}
For $(p_n,\cdots,p_1)\in\{0,1\}^n$ and $(j_n,\cdots,j_1)$
which satisfy the condition
\begin{eqnarray}
&&
j_{l+1}-j_l\geq H(p_{l+1},p_l)
\quad 1\leq l\leq n-1
\label{condi}
\end{eqnarray}
define
\begin{eqnarray}
&&
\cb^{p_n,\cdots,p_1}_{j_n,\cdots,j_1}=
\{
\v^{\ast p_n}_{2j_n+i_n}\cdots\v^{\ast p_1}_{2j_1+i_1}
\in B(p_n,\cdots,p_1)|i_1,\cdots,i_n=0,1
\}.
\nonumber
\end{eqnarray}
Here we set $(j_n,\cdots,j_1)=\phi$ if $n=0$, and $\cb^{\phi}_{\phi}=\{1\}$.
\end{definition}

\noindent
Note that the set
$\cb^{p_n,\cdots,p_1}_{j_n,\cdots,j_1}\sqcup\{0\}$
is preserved by the actions of $\et_1$ and $\ft_1$
and all nonzero elements of it has the same weight with respect to
$d$.
As a corollary of Theorem \ref{base} we have

\begin{theorem}\label{yangian}
There is an isomorphism of $({\widehat{sl_2}})_1$ crystal
\begin{eqnarray}
&&
B(\la_l)\simeq
\sqcup_{n=0}^\infty
\sqcup_{(p_n,\cdots,p_1)\in\pek_{res,n}(0,l)}
\sqcup_{(j_n,\cdots,j_1)}
\cb^{p_n,\cdots,p_1}_{j_n,\cdots,j_1},
\nonumber
\end{eqnarray}
where the disjoint union in $(j_n,\cdots,j_1)$ is over all set
which satisfies the condition (\ref{condi}) and $j_1\geq0$.
\end{theorem}
\vskip3truemm

\noindent
Let us determine the structure of
$\cb^{p_n,\cdots,p_1}_{j_n,\cdots,j_1}$ as a
$({\widehat{sl_2}})_1$ crystal.
The crucial point for it is that an element of the form
$\v^{\ast p_n}_{2j_n+i_n}\cdots\v^{\ast p_1}_{2j_1+i_1}$
can be zero.
We first study the most degenerate cases.

\begin{prop}\label{degen}
If $j_l-j_{l-1}=H(p_l,p_{l-1})$ for any $l\geq2$, then
\begin{eqnarray}
&&
\cb^{p_n,\cdots,p_1}_{j_n,\cdots,j_1}
\simeq
B^{(n)}
\nonumber
\end{eqnarray}
as a $({\widehat{sl_2}})_1$ crystal.
\end{prop}
\par
\noindent
Proof
\par
\noindent
Using
$$
\v^{\ast p_l}_{2j_l}\v^{\ast p_{l-1}}_{2j_{l-1}+1}=0
$$
and the definition of $\ft_1$, $\et_1$, the statement
is easily verified.
$\Box$
\vskip2truemm

\noindent
In order state the general case we need to introduce some
terminology.

\begin{definition}
For $(p_n,\cdots,p_1)\in\{0,1\}^n$ the $(j_n,\cdots,j_1)\in\bz^n$
is said to be a string if
\begin{eqnarray}
&&
j_l-j_{l-1}=H(p_l,p_{l-1})
\quad 2\leq l\leq n.
\end{eqnarray}
We call $n$ the length of the string.
\end{definition}

For a fixed $(p_n,\cdots,p_1)$ we can uniquely decompose
$(j_n,\cdots,j_1)$ into strings in such a way
that
\begin{description}
\item[(i)]
there exist $M\in\bz_{\geq1}$ and $0=r_0<\cdots<r_M=n$
such that $(j_{r_s},\cdots,j_{r_{s-1}+1})$ is a string
for any $s=1,\cdots,M$,
\item[(ii)]
$j_{r_s+1}-j_{r_s}>H(p_{r_s+1},p_{r_s})$ for any $s=1,\cdots,M-1$.
\end{description}
Setting $m_s=r_s-r_{s-1}$, we call this decomposition
the string decomposition of type $(m_M,\cdots,m_1)$.

\begin{theorem}\label{stringd}
Let us fix $(p_n,\cdots,p_1)$.
If the string decomposition of $(j_n,\cdots,j_1)$
is of type $(m_M,\cdots,m_1)$, then
there is an isomorphism
\begin{eqnarray}
&&
\cb^{p_n,\cdots,p_1}_{j_n,\cdots,j_1}
\simeq
B^{(m_M)}\ot\cdots\ot B^{(m_1)}
\nonumber
\end{eqnarray}
of $({\widehat{sl_2}})_1$ crystals.
\end{theorem}
\par
\noindent
Proof
\par
\noindent
Let us use the notations of the above definition
for the string decomosition of $(j_n,\cdots,j_1)$ such as
$r_s$. Set
\begin{eqnarray}
&&
\cb=\{
\v^{\ast p_n}_{2j_n+i_n}\ot\cdots\ot\v^{\ast p_1}_{2j_1+i_1}|
i_n,\cdots,i_1=0,1
\}.
\nonumber
\end{eqnarray}
By definition the product map
$\cb\longrightarrow\cb^{p_n,\cdots,p_1}_{j_n,\cdots,j_1}\sqcup\{0\}$
is factorized as
$$
\begin{array}{ccc}
\cb\sqcup\{0\}&\longrightarrow&\cb^{p_n,\cdots,p_1}_{j_n,\cdots,j_1}
\sqcup\{0\}
\\
\downarrow& &\parallel
\\
\cb^{p_n,\cdots,p_{r_M+1}}_{j_n,\cdots,j_{r_M+1}}
\ot\cdots\ot
\cb^{p_{r_1},\cdots,p_1}_{j_{r_1},\cdots,j_1}\sqcup\{0\}
&\stackrel{F}{\longrightarrow}&
\cb^{p_n,\cdots,p_1}_{j_n,\cdots,j_1}\sqcup\{0\}.
\end{array}
$$
By the definition of the crystal structure
on $\cb^{p_n,\cdots,p_1}_{j_n,\cdots,j_1}$, each map of the
above diagram commutes with $\et_1$ and $\ft_1$.
It is easily verified that $F$ is a bijection.
Then the theorem follows from Proposition \ref{degen}.
$\Box$
\vskip3truemm

To each
$\cb^{p_n,\cdots,p_1}_{j_n,\cdots,j_1}
\simeq
B^{(m_M)}\ot\cdots\ot B^{(m_1)}$, let us define the
$U_q({\widehat{sl_2}})_1$ module $V^{p_n,\cdots,p_1}_{j_n,\cdots,j_1}$
by
$$
V^{p_n,\cdots,p_1}_{j_n,\cdots,j_1}=
V^{(m_M)}\ot\cdots\ot V^{(m_1)},
$$
where $U_q({\widehat{sl_2}})_1$ is the subalgebra generated by
$e_1,f_1,h_1$.
Then, by the uniqueness of the crystal base (Theorem 3 of \cite{K1}),
as a corollary of Theorem \ref{yangian} and Theorem \ref{stringd}
we have

\begin{cor}
Suppose that
$q\in\{q\in{\bf C}|q^n\neq 1,\,n=1,2,\cdots \}\sqcup \{1\}$.
Then there is an isomorphism of $U_q({\widehat{sl_2}})_1$ modules
\begin{eqnarray}
&&
V(\la_l)\simeq
\oplus_{n=0}^\infty
\oplus_{(p_n,\cdots,p_1)\in\pek_{res,n}(0,l)}
\oplus_{(j_n,\cdots,j_1)}
V^{p_n,\cdots,p_1}_{j_n,\cdots,j_1}.
\nonumber
\end{eqnarray}
\end{cor}
\noindent
So far we do not have a natural full Yangian type operations
on  each set $\cb^{p_{n},\cdots,p_1}_{j_{n},\cdots,j_1}$ or
on the $U_q({\widehat{sl_2}})_1$ module
$V^{p_n,\cdots,p_1}_{j_n,\cdots,j_1}$.
One reason of this will be explained in the following way.
We consider not a ${\widehat{sl_2}}$ module but a $U_q({\widehat{sl_2}})$
module. The full Yangian action may not survive
in the deformation but it is possible for the $sl_2$ structure to
outlive. If we see apropriately we can extract the full Yangian
structure from it.

%
%
\section{Character formulas}
Here, we derive the character formulas for the level $k$ integrable
${\widehat {sl_2}}$-modules
$V(\la_{j})$ ($\la_{j}=(k-j)\La_0+j\La_1$, $0 \leq j \leq k$)
using the description of the crystaline spinon basis in section 3 and 4.

Let us denote
$(p_n,\cdots,p_1)\in\{0,1\}^{n}$
the level $k$ restricted path
$(m_n,\cdots,m_0)$
begining $m_0=0$ ending $m_n=j$ by making
identification as $m_l=m_{l-1}+1-2p_l$.

In section 3 (corollary3),
we obtained a spinon parametrization of the basis of $B(\la_{j})$
$$
\v_{j_n}^{\ast p_n}\cdots\v_{j_{1}}^{\ast p_{1}},
\quad n\geq0,
$$
where the conditions for the indices are
\begin{eqnarray}
&j_l\geq j_{l-1}+2  &\quad \hbox{ if $(p_{l},p_{l-1})=(1,0)$},
\nonumber \\
&j_l\geq j_{l-1}    &\quad \hbox{ otherwise},
\nonumber \\
&j_{1}\geq 0.
\nonumber
\end{eqnarray}
The weight $wt$ of which is given by\footnote{
In conformal field theory language, $L_0=-d+\Delta(j)$,
($\Delta(j)=j(j+2)/4(k+2)$)
is the conformal weight and $J^0_0=h_1/2$ is the $U(1)$ charge.}
\begin{eqnarray}
&&-d \equiv \langle -d, wt \rangle=\sum_{l=1}^{n} n_l, \nonumber \\
&&h_1 \equiv \langle h_1, wt \rangle=
\sum_{l=1}^{n} (1-2 r_l)=n-2 m(r), \nonumber
\end{eqnarray}
where $j_l=2 n_l+r_l$, ($r_l \in \{0,1\} $),
and $m(r)=\sharp \{l \ \vert \ r_l=1 \}$.

For fixed $p=(p_n,\cdots,p_{1})$ and
$r=(r_n,\cdots,r_1)$,
the above conditions on $(j_n,\cdots,j_1)$
can be rewritten in terms of
mode sequences $(n_n,\cdots,n_1)$ as
$$
n_l +r_l \geq n_{l-1}+r_{l-1}
+H(r_{l},r_{l-1})+H(p_{l},p_{l-1}), \ (1 \leq l \leq n)
$$
where $r_{0}=p_{0}=0$
and the $H$-function is
$$
H(0,0)=H(0,1)=H(1,1)=0,
\quad
H(1,0)=1.
$$
Especially, the minimal allowed mode sequence is
$$
n_{{\rm min},l}=\sum_{i=1}^{l} (H(r_{i},r_{i-1})+H(p_{i},p_{i-1}))-r_l.
$$
The general mode sequence can be parametrized as
$$
n_l=n_{{\rm min},l}+i_1+\cdots +i_l, \ (i_l \geq 0),
$$
and weight $-d$ of whi{\rm ch} is given by
$$
\sum_{l=1}^n n_l=\sum_{l=1}^{n} l i_l+
\sum_{l=1}^{n} [(n-l+1) H(r_{l},r_{l-1})+(n-l+1) H(p_{l},p_{l-1})-r_l].
$$

In summary, the crystaline spinon basis
can be parametrized by three sequences
$i=(i_n,\cdots,i_1) \in (\bz_{\geq 0})^n$,
$p=(p_n,\cdots,p_{1}) \in \{0,1\}^n$ (restricted) and
$r=(r_n,\cdots,r_1) \in \{0,1\}^n$,
with weights
\begin{eqnarray}
&&-d=h(i)+h(p)+h(r)-m(r), \nonumber \\
&&h_1=n-2 m(r), \nonumber \\
&&h(i)=\sum_{l=1}^{n} l i_l, \nonumber \\
&&h(p)=\sum_{l=1}^{n} (n-l+1) H(p_{l},p_{l-1}), \nonumber \\
&&h(r)=\sum_{l=1}^{n} (n-l+1) H(r_{l},r_{l-1}), \nonumber
\end{eqnarray}

Using this description of basis,
one obtain the following character formula.

\begin{theorem}
The caracter
${\rm ch}_j(q,z) \equiv {\rm tr}_{V(\la_{j})} (q^{-d} z^{h_1})$
is given by
\begin{eqnarray}
{\rm ch}_j(q,z)=\sum_{n=0}^\infty \sum_{m=0}^{n}
{1 \over (q)_{n-m} (q)_{m}}
\sum_{p} q^{h(p)} z^{n-2 m},
\label{character}
\end{eqnarray}
where $p=(p_n,\cdots,p_{1})$ is level $k$ restricted fusion path
from $m_0=0$ to $m_n=j$.
\end{theorem}

Proof.

Using the parametrization of the crystaline basis as above,
one obtain
\begin{eqnarray}
{\rm ch}_j(q,z)
&=&\sum_{n=0}^\infty {1 \over (q)_n}
\sum_{p} \sum_{r}
q^{h(p)+h(r)} q^{-m(r)} z^{n-2 m(r)},
\nonumber \\
&=&\sum_{n=0}^\infty \sum_{m=0}^{n}
{1 \over (q)_{n-m} (q)_{m}}
\sum_{p} q^{h(p)} z^{n-2 m},
\nonumber
\end{eqnarray}
where we have used the formula
$$
\sum_{i_l \geq 0} q^{i_1+2 i_2+\cdots +n i_n}={1 \over (q)_n},
$$
and
$$
\sum_{r=(r_n,\cdots,r_1)} q^{h(r)} q^{-m(r)} z^{-2 m(r)}
=\sum_{m=0}^{n} {(q)_n \over (q)_{n-m} (q)_m} z^{-2 m}.
$$
Thus the formula is proved. $\Box$

This result is exactly the spinon character formula proposed in
\cite{BLS2}.

In the paper \cite{BLS2}, the remaining sum over the restricted fusion path $p$
has also been done explicitly. We quote their final result
$$
{\rm ch}_j(q,z)=q^{-j/4} \sum_{n=0}^\infty \sum_{m=0}^{n}
{1 \over (q)_{n-m} (q)_{m}} z^{n-2 m} q^{-n^2/4} \Phi^n_{A_k} (u_j,q),
$$
and
$$
\Phi^{m_1}_{K} (u,q)=\sum_{m_2, \cdots, m_k} q^{{1 \over 4} m \cdot K \cdot m}
\prod_{i \geq 2}
\Big[ \begin{array}{c} {1 \over 2} ( (2-K) \cdot m +u)_i \\
m_i
\end{array}
\Big].
$$
Here $K$ is the Cartan matrix of $A_k$-type, $(u_j)_i=\delta_{i,j+1}$
and
$$
\Big[ \begin{array}{c} n \\ m
\end{array} \Big]={(q)_n  \over (q)_{n-m} (q)_m}.
$$
The $m$ sum is taken over all non-negative odd [resp. even] integers
for $(m_{j},m_{j-2},\cdots)$ [resp. otherwise].

\section{Quasi-particles at $q=0$}
\par
In this section, as another application of
the crystaline creation algebra, we derive the conjectural formula
for the multi quasi-particle states of the higher spin XXZ models at $q=0$.
In the case of spin $1/2$ XXZ chain,
the formulas are consistent with the results of
the Bethe Ansatz calculations\cite{DFJMN}.

\begin{prop}\label{limit}
For $(p_n,\cdots,p_1)\in\pek_{res,n}$ and
$(\ep_n,\cdots,\ep_1)\in\{0,1\}^n$ we have
\begin{eqnarray}
&&
\v^{\ast p_n}_{\ep_n}(z_n)
\cdots
\v^{\ast p_1}_{\ep_1}(z_1)[[ \ ]]
=
\sum_{(m_1,\cdots,m_n) \in M}C(m_n\cdots m_1)
[[2m_n+\ep_n-p_n,\cdots,2m_1+\ep_1]],
\nonumber
\end{eqnarray}
where
$$
C(m_n\cdots m_1)
=
\prod_{l=1}^n
z_l^{\nu_l}
\det(z_i^{m_j-\nu_j})_{1\leq i,j\leq n},
$$
and
\begin{eqnarray}
\nu_j
=
\sum_{l=j}^{n-1}
\big(
-K(p_{l+1},p_{l})+H(\ep_{l},\ep_{l+1})
\big)
\nonumber
\end{eqnarray}
The sum in $(m_1,\cdots,m_n)$ is over the modes $M$ such as
\begin{eqnarray}
M=\{(m_1,\cdots,m_n) \ \vert \  2m_n+\ep_n-2I_n\geq\cdots\geq 2m_1+\ep_1 \}.
\nonumber
\end{eqnarray}
Here $I_n=I_n(p_n,\cdots,p_1)$ etc.
\end{prop}
\par
\noindent
Proof
\par
\noindent
Let us use the notations in the proof of Theorem \ref{linear}.
Namely, for $\mu=0,1$,
$$
\v^{\ast p}_{2n+\mu}=\psi_{(p,\mu)}(n).
$$
Let us set $A_j=(p_{n-j+1},\ep_{n-j+1})$.
Then, using Lemma \ref{permu}, we have
\begin{eqnarray}
&&
\v^{\ast p_n}_{\ep_n}(z_n)
\cdots
\v^{\ast p_1}_{\ep_1}(z_1)
[[ \ ]]
\nonumber
\\
&=&
\sum_{m_1,\cdots,m_n\in\bz}
\psi_{A_1}(m_1)\cdots \psi_{A_n}(m_n)[[ \ ]]
z_n^{m_1}\cdots z_1^{m_n}
\nonumber
\\
&=&
\sum_{(m_1,\cdots,m_n) \in M}
\sum_{\sigma\in S_n}
{\rm sgn}(\sigma)
z_n^{m_{1,\sigma(1)}}\cdots z_1^{m_{n,\sigma(n)}}
\psi_{A_1}(m_1)\cdots \psi_{A_n}(m_n)[[ \ ]]
\nonumber
\\
&=&
\sum_{(m_1,\cdots,m_n) \in M}
\det(z_{n-i+1}^{s_{i,j}+m_{n-j+1}})
[[2m_n+\ep_n-p_n,\cdots,2m_1+\ep_1]].
\nonumber
\end{eqnarray}
Noting that
$$
\det(z_{n-i+1}^{s_{i,j}+m_{n-j+1}})=
\det(z_i^{m_j+s_{n-i+1,n-j+1}})=
\prod_{l=1}^nz_l^{s_{n-l+1,1}}det(z_i^{m_j+s_{1,n-j+1}})
$$
and $\nu_j=s_{n-j+1,1}=-s_{1,n-j+1}$ we obtain the desired formula.
$\Box$

Let $\{G(b)|b\in B(\la_l)\ot B(\la_0)^\ast\}$ be the
global crystal base of $V(\la_l)\ot V(\la_0)^\ast$\cite{K3} and
$\{ \langle G(b)|\}$ the dual base,
$ \langle G(b) | G(b') \rangle =\delta_{bb'}$.
Let $\v_{\la,\ep}^{\ast\mu}(z)$ be the creation operator
defined in \cite{IIJMNT} acting on the actual representation.
We denote by $| {\rm vac} \rangle_0$ the ground state
in $V(\la_0)\hat{\ot} V(\la_0)^\ast$\cite{IIJMNT}.
Then

\begin{conj}
Assume the same data as in Proposition \ref{limit}.
Let $b=[[2m_n+\ep_n-p_n,\cdots,2m_1+\ep_1]]$ be an element
of $\sqcup_{l=0}^k B(\la_l)\ot B(\la_0)$. Then
the n-quasi-particles have the well defined $q\ra0$ limit and
\begin{eqnarray}
&&
\langle G(b) |
\v^{\ast \la_{i_n}}_{\la_{i_{n-1}}}(z_n)
\cdots
\v^{\ast \la_{i_1}}_{\la_{i_0}}(z_1)
 | {\rm vac} \rangle_0|_{q\ra0}=
\prod_{j=1}^nz_j^{r_j}
C(m_n,\cdots,m_1),
\nonumber
\end{eqnarray}
where $i_0=0$, $i_l-i_{l-1}=1-2p_l$ and
$r_j=\Delta(i_j)-\Delta(i_{j-1})$.
\end{conj}

\section{Discussion}
\par
\noindent
In this paper we have given a spinon type description
of the crystal of level $k$ integrable highest or lowest
weight $\uq$ modules.
The algebra of creation operators for higher spin XXZ model at $q=0$
has been introduced and it plays a central role.
As consequences of this description we derived the fermionic
type character formulas and found the Yangian like structure
in the integrable highest weight modules or in the level
zero modules.
Moreover this description clarifies
the structure of the crystal associated with the tensor products
of the integrable highest and lowest weight modules of the same level
in a different manner from that in \cite{IIJMNT}.
In the following part we shall discuss the possible generalizations and
remaining problems.

To generalize our formalism to the case of arbitrary
affine quantum algebras will be possible.
The problems there are
the identification of creation operators with type II VOs
and to calculate the commuation relations of them.
As to the first problem Reshetikhin\cite{R} gave a conjecture
that the elementary excitations of the model
defined by the affine Lie algebra $\hat{{\cal G}}$
will be parametrized by
the fundamental weights of ${\cal G}$.
We remark that the second process will be skipped
with the aid of the general theory of crystals\cite{KKMMNN1,DJO}.
The treatment of the RSOS models will also be possible\cite{JMO}.

So far we could not prove the existence of the $q\ra0$ limit
of the creation operators. However it is reasonable and certain
that this limit actually exists (see for example \cite{M}).
If the limit exists, the corresponding components in the products
of q-vertex operators to our crystaline spinon basis form a base
of the $\uq$ module.
To understand the "Yangian" representation
$V^{p_n,\cdots,p_1}_{j_n,\cdots,j_1}$ in the integrable
highest weight modules
the study of these problems will be important.
\vskip5truemm

\noindent
{\bf Note}
\par
\noindent
While preparing this munuscript we came to
know that Nakanishi et al.\cite{ANOT} also proved the
character formula which is equivalent to one in section 5
rigorously based on the path description of the character.
They also discuss the Yangian structure from the combinatorial
point of view which is closely related to our results in section 4.
We would like to thank T.Nakanishi for sending us his private
note.

\appendix
\section{Appendix}
Here we consider the naive limit of the $q$-vertex operators
at $q=0$. The consideration here motivates the definition of section 2,
though it tells nothing about the existence.

In the vertex operator formalism, the creation operators are
represented as the type II vertex operator $\v^{\ast \nu}_{\mu \ \ep}(z)$.
The commutation relation of them are given by
$$
\v^{\ast \nu}_{\mu \ \ep_1}(z_1) \v^{\ast \mu}_{\la \ \ep_2}(z_2)=
{R_{V^{\ast} V^{\ast}}}^{\ep'_1 \ \ep'_2}_{\ep_1 \ \ep_2} ({z_1 \over z_2})
\v^{\ast \nu}_{\mu' \ \ep'_2}(z_2) \v^{\ast \mu'}_{\la \ \ep'_1}(z_1)
W( \begin{array}{cc} \la & \mu \\ \mu' & \nu \end{array} \vert {z_1 \over
z_2}).
$$
For the definitions and notations see \cite{DFJMN,IIJMNT,JMO}.
$R_{V^{\ast} V^{\ast}}$ is the trigonometric
$R$-matrix of $U_q({\widehat {sl(2)}})$
for 2-dimensional representation $V^{\ast}$.
$W$ is essentially the elliptic Boltzmann weight due to \cite{ABF}.
Infinitly many poles in $W$ corresponds to the multivaluedness of
usual ($q=1$) CFT chiral vertex operators \cite{TK}.
\footnote{One can take another $q \rightarrow 1$ limit in which
$R$ and $W$ reduce to rational and trigonometric $R$ respectively
by rescaling the spectral parameter $z=q^{-2\beta /i \pi}$.
This limit is exactly what we expected for the "particles",
that is (nilpotent half of) the Faddeev-Zamolodchikov algebra.}

Being regarded as a commutation relation, it looks too terrible to treat.
For instance when $q=1$, the algebra brings some generalized commutation
relation in the sence of the vertex operator algebra which takes very
complicated form in general.
However, if one consider the opposite limit $q \rightarrow 0$,
remarkable simple structure comes out.
In fact one can show that
$$
R_{V^{\ast} V^{\ast}}(z) \rightarrow
-z^{-1/2} \left[
\begin{array}{cccc}1&&& \\ &&z& \\ &1&& \\ &&&1 \end{array}
\right],
$$
$$
W( \begin{array}{cc} \la & \la_{\pm} \\ \la_{\pm} & \la \end{array} \vert z)
\rightarrow z^{\pm 1/2},
$$
$$
W( \begin{array}{cc} \la & \la_{\pm} \\ \la_{\pm} & \la_{\pm \pm} \end{array}
\vert z)
\rightarrow z^{-1/2},
$$
and other $W$'s vanish.
Note that the $W$ are independent of the initial weight $\la$ in this limit,
hence, one can regard it as a vertex type weight.
Accordingly, we change the notation as
$$
\v^{\ast \la_{\pm}}_{\la \ \ep}(z) \rightarrow \v^{\ast \pm}_{\ep} (z),
$$
then the commutation relation takes the form
$$
\v^{\ast p_1}_{\ep_1}(z_1) \v^{\ast p_2}_{\ep_2}(z_2)=
-({z_1 \over z_2})^e \v^{\ast p_1}_{\ep_1}(z_2) \v^{\ast p_2}_{\ep_2}(z_1),
$$
where the exponent $e$ is given by
$$
e=H(p_1, p_2)+H(\ep_2, \ep_1)-1.
$$
This algebra can be regarded as a twisted version of anti-commutation
relation.
The algebra in the main text (Definition 1) can be derived by
mode expansion as follows
\begin{eqnarray}
&&\v^{\ast p}_0 (z)=\sum_{n \in \bz} \v^{\ast p}_{2 n} z^n, \nonumber \\
&&\v^{\ast p}_1 (z)=\sum_{n \in \bz} \v^{\ast p}_{2 n+1} z^n. \nonumber
\end{eqnarray}

\section{Appendix}
\begin{table}
\caption{crystaline spinon basis of $V(\Lambda_0)$}
\begin{tabular}{|c|c|c|c|c|}
\hline
$(-d,h_1)$ & crystal base & [[domain wall]] & (fusion path) & [$j$-path] \\
\hline
(0,0) & $\cdots$ 010101 & [[\ ]] & (\ ) & [\ ] \\
\hline
(1,2) &  $\cdots$ 01010/0/ & [[1,0]] & (1,0) & [2;0] \\
(1,0) &  $\cdots$ 0101/10/ & [[2,0]] & (1,0) & [3;0] \\
(1,-2) & $\cdots$ 0101/1/1 & [[2,1]] & (1,0) & [3;1] \\
\hline
(2,2) & $\cdots$ 010/010/ & [[3,0]] & (1,0) & [4;0] \\
(2,0) & $\cdots$ 010/01/1 & [[3,1]] & (1,0) & [4;1] \\
(2,0) & $\cdots$ 01/1010/ & [[4,0]] & (1,0) & [5;0] \\
(2,-2) & $\cdots$ 01/101/1 & [[4,1]] & (1,0) & [5;1] \\
\hline
$\cdots$ \\
\hline
(4,4) & $\cdots$ 010/0/0/0/ & [[3,2,1,0]] & (1,0,1,0) & [4;2,2;0] \\
\hline
\end{tabular}
\end{table}

\begin{table}
\caption{crystaline spinon basis of $V(\Lambda_1)$}
\begin{tabular}{|c|c|c|c|c|}
\hline
$(-d,h_1)$ & crystal base & [[domain wall]] & (fusion path) & [$j$-path] \\
\hline
(0,1)  & $\cdots$ 101010/ & [[0]] & (0) & [0] \\
(0,-1) & $\cdots$ 10101/1 & [[1]] & (0) & [1] \\
\hline
(1,1) &   $\cdots$ 1010/01 & [[2]] & (0) & [2] \\
(1,-1) &  $\cdots$ 101/101 & [[3]] & (0) & [3] \\
\hline
(2,3)  & $\cdots$ 1010/0/0/ & [[2,1,0]] & (0,1,0) & [2,2;0] \\
(2,1)  & $\cdots$ 10/0101 & [[4]] & (0) & [4] \\
(2,1)  & $\cdots$ 101/10/0/ & [[3,1,0]] & (0,1,0) & [3,2;0] \\
(2,-1) & $\cdots$ 1/10101 & [[5]] & (0) & [5] \\
(2,-1) & $\cdots$ 101/1/10/ & [[3,2,0]] & (0,1,0) & [3,3;0] \\
(2,-3) & $\cdots$ 101/1/1/1 & [[3,2,1]] & (0,1,0) & [3,3;1] \\
\hline
\end{tabular}
\end{table}


\begin{table}
\caption{crystaline spinon basis of $V(2 \Lambda_0)$}
\begin{tabular}{|c|c|c|c|c|}
\hline
$(-d,h_1)$ & crystal base & [[domain wall]] & (fusion path) & [$j$-path] \\
\hline
(0,0) & $\cdots$ 020202 & [[\ ]] & (\ ) & [\ ] \\
\hline
(1,2) & $\cdots$ 02020/1/ & [[1,0]] & (1,0) & [2;0] \\
(1,0) & $\cdots$ 0202/11/ & [[2,0]] & (1,0) & [3;0] \\
(1,-2) & $\cdots$ 0202/1/2 & [[2,1]] & (1,0) & [3;1] \\
\hline
(2,4) & $\cdots$ 02020//0// & [[1,1,0,0]] & (1,1,0,0) & [2,2;0,0] \\
(2,2) & $\cdots$ 020/111/ & [[3,0]] & (1,0) & [4;0] \\
(2,2) & $\cdots$ 0202/1/0// & [[2,1,0,0]] & (1,1,0,0) & [3,2;0,0] \\
(2,0) & $\cdots$ 02/1111/ & [[4,0]] & (1,0) & [5;0] \\
(2,0) & $\cdots$ 0202//20// & [[2,2,0,0]] & (1,1,0,0) & [3,3;0,0] \\
(2,0) & $\cdots$ 020/11/2 & [[3,1]] & (1,0) & [4;1] \\
\hline
\end{tabular}
\end{table}

\begin{table}
\caption{crystaline spinon basis of $V(\Lambda_0+\Lambda_1)$}
\begin{tabular}{|c|c|c|c|c|}
\hline
$(-d,h_1)$ & crystal base & [[domain wall]] & (fusion path) & [$j$-path] \\
\hline
(0,1) & $\cdots$ 111111/ & [[0]] & (0) & [0] \\
(0,-1) & $\cdots$ 11111/2 & [[1]] & (0) & [1] \\
\hline
(1,3) & $\cdots$ 1111/0// & [[1,0,0]] & (1,0,0) & [2;0,0] \\
(1,1) & $\cdots$ 1111/02 & [[2]] & (0) & [2] \\
(1,1) & $\cdots$ 1111/20// & [[2,0,0]] & (1,0,0) & [3;0,0] \\
(1,-1) & $\cdots$ 111/202 & [[3]] & (0) & [3] \\
(1,-1) & $\cdots$ 1111/0/1/ & [[2,1,0]] & (1,0,0) & [3;1,0] \\
(1,-3) & $\cdots$ 1111/2//2 & [[2,1,1]] & (1,0,0) & [3;1,1] \\
\hline
(2,3) & $\cdots$ 111/020// & [[3,0,0]] & (1,0,0) & [4;0,0] \\
(2,3) & $\cdots$ 1111/0/1/ & [[2,1,0]] & (0,1,0) & [2,2;0] \\
\hline
\end{tabular}
\end{table}

\begin{table}
\caption{crystaline spinon basis of $V(2 \Lambda_1)$}
\begin{tabular}{|c|c|c|c|c|}
\hline
$(-d,h_1)$ & crystal base & [[domain wall]] & (fusion path) & [$j$-path] \\
\hline
(0,2) & $\cdots$ 202020// & [[0,0]] & (0,0) & [0,0] \\
(0,0) & $\cdots$ 20202/1/ & [[1,0]] & (0,0) & [1,0] \\
(0,-2) & $\cdots$ 20202//2 & [[1,1]] & (0,0) & [1,1] \\
\hline
(1,2) & $\cdots$ 2020/11/ & [[2,0]] & (0,0) & [2,0] \\
(1,0) & $\cdots$ 202/111/ & [[3,0]] & (0,0) & [3,0] \\
(1,0) & $\cdots$ 2020/1/2 & [[2,1]] & (0,0) & [2,1] \\
(1,-2) & $\cdots$ 202/11/2 & [[3,1]] & (0,0) & [3,1] \\
\hline
$\cdots$ \\
\hline
(3,4) & $\cdots$ 2020//0/1/ & [[2,2,1,0]] & (0,0,1,0) & [2,2,2;0] \\
\hline
\end{tabular}
\end{table}

\end{document}